# Slow-wave effect and mode-profile matching in Photonic Crystal microcavities

**C. Sauvan, P. Lalanne and J.P. Hugonin**

Laboratoire Charles Fabry de l'Institut d'Optique, Centre National de la Recherche Scientifique, F-91403 Orsay cedex, France

**Abstract:**

Physical mechanisms involved in the light confinement in photonic crystal slab microcavities are investigated. We first present a full three-dimensional numerical study of these microcavities. Then, to gain physical insight into the confinement mechanisms, we develop a Fabry-Perot model. This model provides accurate predictions and sheds new light on the physics of light confinement. We clearly identify two mechanisms to enhance the Q factor of these microcavities. The first one consists in improving the mode-profile matching at the cavity terminations and the second one in using a slow wave in the cavity.

PACS :

42.55.Sa Microcavity and microdisk lasers

42.70.Qs Photonic bandgap materials

42.55.Tv Photonic crystal lasers

42.50.Pq Cavity quantum electrodynamics

Electromagnetic resonant cavities, which trap light within a finite volume, are essential components of many important optical devices and effects. Besides standard applications of these structures as lasers or frequency filters, they can potentially be used in solid-state quantum electrodynamics experiments to enhance or reduce the spontaneous emission rate[1] and in related applications to quantum information studies.[2] An ideal cavity would show infinitely narrow resonances at discrete frequencies and would confine light indefinitely. Deviation from this ideal condition is described by the cavity quality factor, a quantity related to the amount of light leaking out of the cavity. Quality factor Q and modal volume V, a quantity related to the degree of the cavity-mode confinement, are the prominent parameters in applications of these devices.[1] At optical frequencies, due to the lack of good metals, impetus for ultrahigh Q research has concentrated during the last decade on small volume resonant dielectric cavities like microposts, microdisks, Photonic Crystal (PC) microcavities, microspheres and ring resonators.[3]

Recently, several numerical and experimental studies[4,5,6,7,8,9] have shown that microcavities in two-dimensional PC slabs are highly valuable candidates for achieving high quality factors with wavelength-sized modal volumes. All these works evidence the importance of finely tuning the holes position at the cavity termination. For instance in Ref. 5, a remarkable control over fabricated geometries has been demonstrated, and Q factors of 45,000 have been measured in a silicon-based two-dimensional PC microcavity. The very high Q value, 10 times larger than in previous studies with comparable ultra-small volumes close to $(\lambda/n)^3$, has been achieved through a surprising 10-times Q enhancement induced by a fine shift (60 nm) of the holes surrounding the defect region. In order to estimate the ultimate potential of these microcavities for future applications, it appears essential to understand in depth the role of the hole tuning in the experiment. A good understanding is also important for new designs of microcavities in general.

In this work, we study PC slab microcavities which consist in a small line-defect composed of N missing holes, see Fig. 1 for a schematic view of the structure along with a definition of the different parameters. We first perform a full electromagnetic study of these microcavities with a 3D frequency-domain modal method[10] and emphasize the impact of hole tuning on the Q factors of cavities with several N values. Currently, the theoretical analysis[5,11,12] of PC slab microcavities relies on a complete resolution of the electromagnetic problem with 3D Finite-Difference-Time-Domain methods, followed by an analysis of the cavity-mode pattern through a Fourier (or momentum-space) decomposition method. The approach we develop here is radically different. In comparison with the analysis performed in Ref. 5 which is presently being debated,[13] it also leads to a different interpretation of the physics of light confinement. Instead of looking at a global property of the cavity, we emphasize mirror properties through a Fabry-Perot (FP) model. Interpreting light confinement with a FP model is the prime natural approach one has in mind to analyse in-line PC microcavities,[14,15] but to our knowledge, this work is the first one to validate such an approach for 2D PC microcavities. New physical effects are highlighted, namely an improved mode-profile matching at the cavity terminations and a slow wave effect in the cavity. Similar conclusions have been derived in Ref. 13 but, due to space limitation, they have neither been validated nor been argued in depth.

We first analyse the electromagnetic properties of microcavities for several values of the number N of missing holes. For the calculation, we assume that the refractive index (n = 3.42) of the silicon slab is independent of the frequency, an assumption highly legitimate in the narrow spectral range of interest, and we use a three-dimensional frequency-domain modal method.[10] The cavity Q factors are computed as $Q = \text{Re}(\tilde{\lambda})/2\text{Im}(\tilde{\lambda})$, where $\tilde{\lambda}$ is the complex pole associated to a scattering matrix relating the electromagnetic fields between two outer planes which are parallel to planes P and P' (see Fig. 1) and which are located fifteen rows of holes away from them. By varying this number from 12 to 18, we have checked that the calculated Q's remain unchanged, indicating that in-plane losses are kept at a negligible level in the calculation. Therefore the calculated Q's are intrinsic quality factors solely limited by out-of-plane radiation. The method relies on an analytical integration of Maxwell's equations along the ΓK direction and on a supercell approach in the two others. Periodic boundary conditions are used in the ΓM direction and Perfectly-Matched Layers[16] are used in the z-direction to carefully handle the far-field radiation in the air clads. Since these layers absorb non-evanescent radiations, the electromagnetic fields are null on the z-boundaries of the supercell and are thus periodic functions of the z-coordinates. This allows the calculation of the radiated and guided modes in a Fourier (plane-wave) basis in each layer (the hole shapes are discretized in a series of thin uniform layers) and the integration in the ΓK direction by relating recursively the mode amplitudes in the different layers using a scattering matrix approach.

The calculated Q's (circles) are displayed in Figs. 2(a), 2(b) and 2(c) as a function of the hole displacement $d$. For the sake of comparison, we also include experimental data (squares). These data were determined in Ref. 5 from cavity-radiation-spectrum measurements by removing the coupling effect between the cavity and the probe waveguide. As shown in Fig. 2(c), the calculations quantitatively agree with the experimental data, indicating that inevitable additional losses related to fabrication errors are kept at a rather small level in the experiment. The calculation also predicts an increase of Q by a factor 20 for $d \approx 0.18a$, a value slightly larger than the value of $d = 0.15a$, for which a ≈8-times Q enhancement has been observed experimentally. Comparison of Figs. 2(a), 2(b) and 2(c) shows similar trends for the three geometries: an asymmetric peak with a peak quality factor $Q_{max}$ achieved for $d/a \approx 0.18$. We additionally note that $Q_{max}$ increases rapidly as N increases, starting from 4000 for N = 1 and reaching a value of 92,000 for N = 3. These important characteristics will be explained in the following.

To gain physical insight into the mechanisms of the Q enhancement, we now develop a FP model and consider the PC microcavity as a FP resonator composed of a single-line-defect PC waveguide closed at both extremities by two PC mirrors. Within this approach, the cavity mode results from the bouncing of the fundamental Bloch mode of the PC waveguide between two PC mirrors along the ΓK direction. In the spectral range of interest, this Bloch mode is a truly lossless guided mode operating below the light line of the air-clad.[17,18] Therefore the finite lifetime of the cavity mode is solely due to the imperfect Bloch mode reflectivity $|r(\lambda)|^2$ of the PC mirrors, which is strictly smaller than unity. Since in-plane losses are null, the quantity $\mathcal{L} = 1-|r(\lambda)|^2$ represents inevitable out-of-plane radiation losses occurring when the mode of the PC waveguide is impinging on the mirror. Hereafter, the effective index $n_{eff}(\lambda)$ of the fundamental Bloch mode of the single-line-defect PC waveguide and its modal reflectivity coefficient $|r(\lambda)|\exp[i\varphi_r(\lambda)]$ on the PC mirror defined at plane P will play a central role. The Bloch mode is calculated as the eigenstate of the PC waveguide in the Fourier basis,[19] and its modal reflectivity $r(\lambda)$ is obtained with the method of Ref. 10.

Within the FP model, a resonance at a wavelength $\lambda_0$ is due to a phase-matching condition for the Bloch mode. More precisely, the total phase delay $\Phi_T(\lambda_0)$ experienced by the mode along one-half cavity cycle has to be equal to a multiple of $\pi$[20]

$$\Phi_T(\lambda_0) = (2\pi/\lambda_0)n_{eff}(\lambda_0)L + \varphi_r(\lambda_0) = p\pi, \qquad (1)$$

where L is the cavity length equal to N$a$ (see Fig. 1) and p is an integer. Figure 3 shows the total phase delay for different hole displacements, $d/a = 0, 0.05, 0.1, 0.15$ and $0.25$, and for the cavity with N = 3. As $d$ increases, the total phase delay, or equivalently the effective cavity length, increases. Since the cavity-mode order (p = 4) is the same for all $d$'s, the hole displacement results in a red-shift of the resonant wavelength $\lambda_0$. Figure 2(d) compares the FP predictions for the resonant wavelength $\lambda_0$ with experimental values (squares). Good agreement is obtained. In addition, we also note that the FP predictions nicely agree with numerical data for $\text{Re}(\tilde{\lambda})$ (circles). Similar agreements (not shown for the sake of conciseness) have also been obtained for the cavities with N = 1 and 2, for p = 2 and 3, respectively.

The mode lifetime is related to the cavity quality factor defined by $Q = \lambda_0/\Delta\lambda$, where $\Delta\lambda$ is the resonance width at half maximum. For a FP resonator and under the legitimate assumption of a narrow resonance, $\Delta\lambda$ can be straightforwardly expressed as the derivative of $\Phi_T(\lambda)$. One obtains the following expression for Q:

$$Q = \frac{\pi}{1-|r(\lambda_0)|^2}\left[2\frac{L}{\lambda_0}n_g(\lambda_0) - \frac{\lambda_0}{\pi}\left(\frac{\partial \varphi_r}{\partial \lambda}\right)_{\lambda_0}\right]. \qquad (2)$$

In Eq. (2), $n_g$ represents the group index ($n_g = c/v_g$) of the Bloch mode cycling in the resonator. Although the $n_g$ factor is often omitted and replaced by $n_{\text{eff}}$ in classical textbooks[24] - probably because $n_g$ cannot exceed a few unity for classical waveguides even with large refractive-index contrasts-, it must not be omitted in the present study since PC waveguides potentially offer extremely large group-velocity reduction.[20] Whether one understands the $n_g$ factor in Eq. (2) as an increase of the effective cavity length or as an increase of the photon density inside the resonator, it is worthy to mention that $n_g$ mainly affects the cavity-mode lifetime. Thus $n_g$ represents a highly valuable parameter for cavity designs in quantum electrodynamics experiments. Equation (2) highlights also two other physical quantities which impact the mode lifetime: the derivative of the phase of the modal reflectivity coefficient (usually a negative quantity), and the modal reflectivity $|r(\lambda)|^2$ of the Bloch mode, a quantity directly related to the out-of-plane radiation losses $\mathcal{L} = 1 - |r(\lambda)|^2$ incurred at the cavity

terminations. The effects of the hole shift on these three quantities are shown in Figs. 4(a), 4(b) and 4(c) for N = 3. From these calculations and by use of Eq. (2), the FP predictions for the Q factor are derived. As shown in Fig. 2(c), these predictions well agree with experimental data (squares) and with numerical data (circles) obtained by the pole calculation. We have also performed similar comparisons for the cavities with N = 1 and 2 missing holes and, as shown in Figs. 2(a) and 2(b), a quantitative agreement is again achieved.

In general, one does not expect perfect agreement between the FP model predictions and calculation data obtained with the pole approach. The reason is that, in the FP approach, energy transport in between planes P and P' is solely ensured by the fundamental propagating Bloch mode of the PC waveguide, all other energy-transport routes being neglected. These other routes (which are indeed taken into account with the pole calculation) are all the other Bloch modes of the PC waveguide. These modes are all leaky in the spectral range of interest and their leakage guarantees that their impact on the cavity mode lifetime vanishes as N increases. However, a leakage does not preclude an impact by evanescent coupling in ultra-small cavities. Indeed, it has been demonstrated that energy transportation through leaky waves is not negligible under specific conditions and can even drastically impacts the performance of ultra-small air-bridge microcavities.[15] However, the good agreement shown in Figs. 2(a)-(c) suggests that this impact can be neglected in the 2D geometry considered here. Therefore the single Bloch-mode FP model can be used with confidence to analyse the electromagnetic properties of ultra-small PC microcavities even with only one missing hole. This result opens interesting perspectives for future designs.

With the FP model, we are now able to interpret the experimental observation of a Q enhancement by holes tuning in Ref. 5. The first mechanism is a progressive increase of the reflectivity $|r|^2$ [see Fig. 4(c)]. This increase is followed by a quick drop for $d/a > 0.18$, which is responsible for the observation of an asymmetric peak for Q. In Fig. 4(d), we compare $|r(\lambda)|^2$ for two mirrors with $d = 0$ and $d = 0.18a$. The spectrum covers the whole domain of resonance frequencies of the three cavities. General trends are an increase of the reflectivity with the wavelength and an additional increase by shifting the holes, $\mathcal{L}$ being roughly reduced by a constant factor of ≈5 over the entire spectrum. The reflectivity increase induced by hole tuning has been previously interpreted for one-dimensional Bragg reflectors[14,15] and more recently for two-dimensional PC mirrors.[21] This effect is understood as a mode conversion in the non-periodic region of the mirror: the hole tuning decreases the mode-profile mismatch between the propagative Bloch mode of the PC waveguide and the evanescent Bloch mode of the PC mirror. The accuracy of the FP model predictions in Figs. 2(a)-(c) evidences the strong relationship between the mode lifetime of a two-dimensional PC microcavity and mode-profile matching problems at the cavity terminations.

The second mechanism involved in the Q enhancement is an increase of the group index $n_g$ and of $\partial\varphi_r/\partial\lambda$ [see Figs. 4(a)-(b)]. This increase results from the highly dispersive nature of the Bloch mode, as shown by the dispersion diagram in the inset of Fig. 3, and from the slightly longer penetration depth into the PC mirrors due to the broken periodicity and to the propagation of a slower wave on a longer effective cavity length (L+2$d$). Whereas the reflectivity increase is roughly the same for the three cavities (N = 1, N = 2 and N = 3), the slow-wave effect strongly depends on the cavity length. For N = 3 missing holes, the group velocity of the Bloch mode is roughly decreased by a factor 2 by shifting the holes while it is roughly unchanged for the cavity with N = 1. To our knowledge, the work in Ref. 5 represents the first demonstration of the reinforcement of light-matter interaction by use of slow-waves in microcavities, a component whose essence is this reinforcement.

In conclusion, we have provided a thorough electromagnetic analysis of light confinement in two-dimensional PC slab microcavities. A classical FP model has been proposed and validated through comparison with rigorous numerical results and experimental data. This suggests that light confinement in PC slab microcavities can be largely understood as one classically does for more traditional in-line cavities. The model provides an analytical expression for the Q factor and its domain of validity surprisingly extends to ultra-small cavities even with a single missing hole. As general design rules, the model suggests that the cavity terminations have to be carefully handled for lowering radiation losses into the cladding and that the resonance frequency of the cavity should match that of a truly propagative Bloch mode below the light line. In addition, the FP model has been used to interpret the observations[5] of very high Q factors in these cavities by finely tuning the hole geometry at the cavity terminations. Two physical effects which have not been mentioned in the original interpretation and which are important for microcavity designs in general have been highlighted. Firstly, the fine tuning of the holes at the cavity terminations impacts the intrinsic properties of the PC mirrors. A small hole shift results in a better mirror performance, i.e. in a reduction of the radiation in the claddings. Secondly, the tuning results in a red-shift of the resonance wavelength and in a decrease of the group velocity of the Bloch mode cycling inside the cavity. Although kept at a moderate level in the experiment, the group-velocity issue has to be considered for further designs. Confining slow light in small volumes is an important outcome of the present analysis, and PC geometries which offer highly dispersive Bloch modes are certainly promising candidates. For instance and for the geometry considered in this work, the FP model predicts a peak Q factor in excess of 250,000 with group velocities of $c/25$ for a cavity with four missing holes.

**Acknowledgments**

C. Sauvan is grateful to the Délégation Générale pour l'Armement for his PhD fellowship.

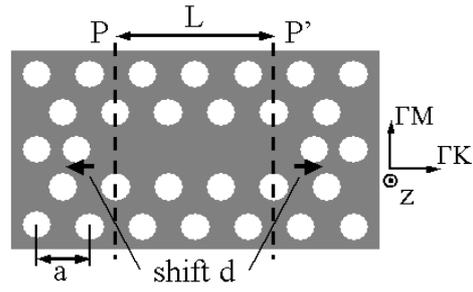

**Figure 1**

Schematic top view of the investigated microcavities formed by filling N holes in the ΓK direction of a two-dimensional PC structure composed of a triangular lattice of air holes (lattice constant a = 0.42 µm) etched into a silicon slab. The picture holds for a cavity with three missing holes (N = 3). The slab thickness is 0.6a and the air holes radii 0.29a. The hole displacement at the cavity edges is denoted by $d$. Planes P and P' are used as phase references for the modal reflectivity used in the Fabry-Perot model.

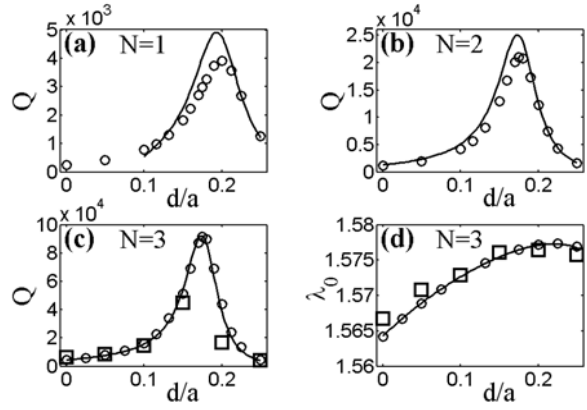

**Figure 2**

Comparison between experimental data (squares), calculation data (circles) and FP model predictions (solid curves). (a), (b) and (c) Q-enhancement for N = 1, 2 and 3, respectively. In (a), the FP model predictions do not show up for $d \leq 0.1a$. For small $d$'s, the resonance wavelength is beneath 1.5 μm and the Bloch mode of the single-line-defect PC waveguide leaks in the air clads. (d) Resonance wavelength red-shift for N = 3.

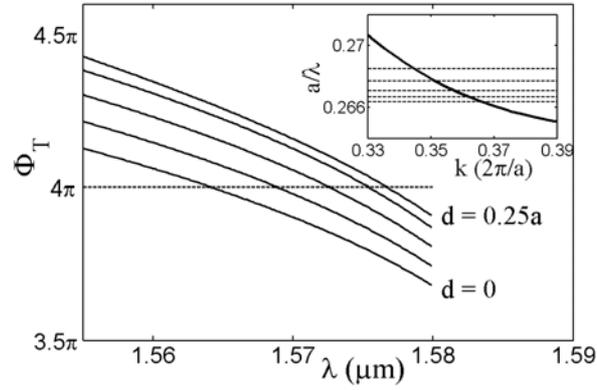

**Figure 3**

The total-phase delay $\Phi_T(\lambda)$ has been represented for $d/a$ = 0, 0.05, 0.1, 0.15 and 0.25 and for N = 3. The phase-matching occurs for $\Phi_T(\lambda_0) = 4\pi$ (dashed horizontal line). Inset: dispersion diagram of the fundamental Bloch mode (solid curve). The dashed lines correspond to the resonance wavelengths for the above-mentioned values of $d$. From top to bottom, $d$ increases.

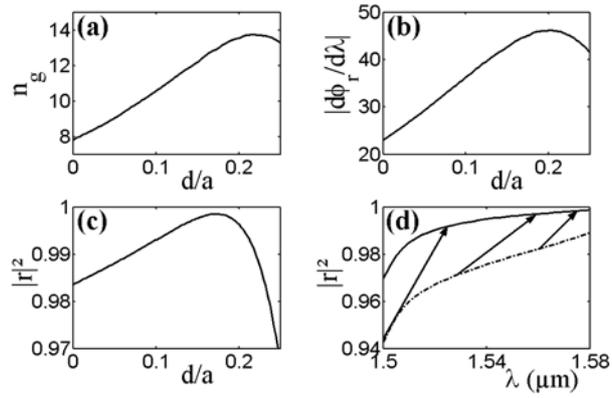

**Figure 4**

(a), (b) and (c) Effect of the holes shift on the main physical parameters involved in Eq. (2) for N = 3. (d) $|r(\lambda)|^2$ for two mirrors with $d = 0$ (dashed-dotted curve) and $d = 0.18a$ (solid curve). The arrows indicate the reflectivity change for the three cavities. From left to right, N = 1, 2 and 3.

# Figure Captions

**Figure 1**

Schematic top view of the investigated microcavities formed by filling N holes in the ΓK direction of a two-dimensional PC structure composed of a triangular lattice of air holes (lattice constant a = 0.42 µm) etched into a silicon slab. The picture holds for a cavity with three missing holes (N = 3). The slab thickness is 0.6a and the air holes radii 0.29a. The hole displacement at the cavity edges is denoted by *d*. Planes P and P' are used as phase references for the modal reflectivity used in the Fabry-Perot model.

**Figure 2**

Comparison between experimental data (squares), calculation data (circles) and FP model predictions (solid curves). (a), (b) and (c) Q-enhancement for N = 1, 2 and 3, respectively. In (a), the FP model predictions do not show up for $d \leq 0.1a$. For small *d*'s, the resonance wavelength is beneath 1.5 µm and the Bloch mode of the single-line-defect PC waveguide leaks in the air clads. (d) Resonance wavelength red-shift for N = 3.

**Figure 3**

The total-phase delay $\Phi_T(\lambda)$ has been represented for *d/a* = 0, 0.05, 0.1, 0.15 and 0.25 and for N = 3. The phase-matching occurs for $\Phi_T(\lambda_0) = 4\pi$ (dashed horizontal line). Inset: dispersion diagram of the fundamental Bloch mode (solid curve). The dashed lines correspond to the resonance wavelengths for the above-mentioned values of *d*. From top to bottom, *d* increases.

**Figure 4**

(a), (b) and (c) Effect of the holes shift on the main physical parameters involved in Eq. (2) for N = 3. (d) $|r(\lambda)|^2$ for two mirrors with $d = 0$ (dashed-dotted curve) and $d = 0.18a$ (solid curve). The arrows indicate the reflectivity change for the three cavities. From left to right, N = 1, 2 and 3.